\documentclass[aps,preprint,superscriptaddress,nofootinbib]{revtex4}

\usepackage{amsmath}
\usepackage{graphicx,subfigure}
\usepackage{color}
\usepackage{slashed}

\usepackage{amssymb,amsmath,amsthm,ulem}

\pdfoutput=1

\usepackage{epsfig}
\usepackage{amsfonts}
\usepackage[colorlinks,
            linkcolor=black,
            anchorcolor=black,
            citecolor=blue
            ]{hyperref}
\usepackage[a4paper,left=2.cm,right=2.cm,top=2.5cm,bottom=2.5cm]{geometry}

\begin{document}
	
\title{ Weak cosmic censorship with self-interacting scalar and bound on charge to mass ratio }

\author{Yan Song\footnote{songy18@lzu.edu.cn}, Tong-Tong Hu\footnote{hutt17@lzu.edu.cn},
and Yong-Qiang Wang\footnote{ yqwang@lzu.edu.cn, corresponding author}}

\affiliation{ Research Center of Gravitation $\&$
 Institute of Theoretical Physics $\&$ \\
Key Laboratory for Magnetism and Magnetic of the Ministry of Education,\\ Lanzhou University, Lanzhou 730000, China}

\vfill
\vfill

\begin{abstract}
 We study the model of Einstein-Maxwell theory minimally coupling to a massive charged self-interacting scalar field, parameterized by the quartic and hexic coupling, labelled by $\lambda$ and $\beta$, respectively. In the absence of scalar field, there is a class of counterexamples to cosmic censorship. Moveover, we investigate the properties of full nonlinear solution with nonzero scalar field, and argue that, by assuming massive charged self-interacting scalar field with sufficiently large charge above one certain bound, these counterexamples can be removed. In particular, this bound on charge for self-interacting scalar field is no longer equal to the weak gravity bound for free scalar case. In the quartic case, the bounds are below free scalar case for $\lambda<0$, while above free scalar case for $\lambda>0$. Meanwhile, in the hexic case, the bounds are above free scalar case for both $\beta>0$ and $\beta<0$.

\end{abstract}

\maketitle


\newpage
\tableofcontents
\newpage

\section{Introduction}\label{sec1}
The weak cosmic censorship conjecture (WCCC), put forward by Penrose \cite{Penrose:1969pc}, states that there is no naked singularity, a region of arbitrarily large curvature visible to distant observers, and the singularity that arise in the gravitational collapse \cite{Hawking:1976ra, Joshi:2012mk} must be hidden within event horizon, never to be observed by external observers. In past several decades, many significant works based on Wald's gedanken experiment have been done to check this conjecture. Some of them have indeed proved its validity under certain conditions \cite{Wald:1974mk,Sorce:2017dst,Rocha:2011wp,Jiang:2019ige,Ying:2020bch,Wang:2020osg,Ge:2017vun,Chen:2019nhv,Gwak:2018akg,Yang:2020czk,Yang:2020iat,Chen:2018yah,Liang:2018wzd,Hod:2008zza},  while the others attempt to present its violations \cite{Hubeny:1998ga,Jacobson:2009kt,Matsas:2007bj,Matsas:2009ww,Richartz:2011vf}. Besides, more examples to violate cosmic censorship are also proposed, for example, a naked singularity can arise from smooth initial data due to the non-linear instability of a black hole \cite{Lehner:2010pn,Figueras:2015hkb,Figueras:2017zwa},
and the Gregory-Laflamme instability in higher dimensions can also cause the horizon unstable to pinch off in finite time, leading to violation of the cosmic censorship \cite{Gregory:1993vy, Hubeny:2002xn,Santos:2015iua}. In addition, based on the four-dimensional Einstein-Maxwell gravity, a promising class of counterexamples with asymptotically anti de-Sitter (AdS) boundary condition was proposed \cite{Horowitz:2016ezu,Crisford:2017zpi}, and not involving Maxwell field, the authors also constructed a class of vacuum counterexamples with differential rotation boundary in \cite{Crisford:2018qkz}.

Recently, the authors \cite{Crisford:2017gsb} proposed that this class of counterexamples in \cite{Horowitz:2016ezu} can be removed by assuming the weak gravity conjecture (WGC) \cite{ArkaniHamed:2006dz,Palti:2019pca,Palti:2017elp,Saraswat:2016eaz}, which can tell whether or not a low-energy effective theory is consistent with quantum theory and ensure that gravity is the weakest force. As discussed in \cite{Crisford:2017gsb},
adding massive charged scalar field with sufficiently large charge, the original Einstein-Maxwell solution in AdS becomes unstable and evolves into new stationary solution with scalar hair, the dual analogy of holographic superconductor \cite{Hartnoll:2008vx,Cai:2015cya,Hartnoll:2008kx}, such that the cosmic censorship is no longer violated in previous way.
In particular,  the minimum charge-to-mass ratio required to preserve cosmic censorship is precisely the bound predicted by weak gravity conjecture.
It asserts that if the weak gravity conjecture holds, cosmic censorship can be preserved. For more evidence, see \cite{Horowitz:2019eum}. Moreover, studying the case of Born-Infeld electrodynamics coupling to gravity \cite{Hu:2019rpw}, the authors also constructed analogous counterexamples to cosmic censorship and found that these examples can still be removed with massive highly charged scalar field.

An important and general way of generalizing free scalar field is to introduce quartic or hexic self-interaction \cite{Herdeiro:2015tia,Herdeiro:2020xmb,Li:2020ffy,Herdeiro:2016gxs,Valdez-Alvarado:2020vqa}, which in general leads to attractive or repulsive force  through the choice of its quartic and hexic coupling parameters. We ask what happens to the bound on charge required to preserve cosmic censorship, if we consider the self-interacting scalar field, instead of non-self-interacting one. To answer this, in this work, we extend the results in \cite{Crisford:2017gsb} and calculate the bounds for various self-interacting couplings.

The paper is organised as follows. In Sec. \ref{sec2}, we construct Einstein-Maxwell-scalar model, taking in account the quartic and hexic self-interaction. In Sec. \ref{sec3}, we present the full nonlinear solutions with charged quartic scalar field and charged hexic scalar field, respectively, and then study their properties, giving the the bound on charge required to preserve cosmic censorship. Finial, we conclude our results in Sec. \ref{sec4}.

\section{The Model and Ansatz}\label{sec2}
To remove the counterexamples proposed in \cite{Horowitz:2016ezu}, we add massive charged self-interacting scalar field $\Phi$ into the original Einstein-Maxwell gravity. The resulting action is given by
\begin{subequations}
\begin{align}
S=\frac{1}{16\pi G}\int d^{4}x\sqrt{-g}[R+\frac{6}{L^{2}}-F^{ab}F_{ab}+\mathcal{L}_{scalar}]\,, \label{eq:einstein}
\end{align}
where $L$ is the AdS length scale for which we take $L=1$, and $G$ the gravitational constant. Moreover, $F_{ab}$ is the electromagnetic strength tensor with respect to $A$, the $U(1)$ gauge potential, by $F\equiv dA$. The scalar term is given by
\begin{align}
&\mathcal{L}_{scalar}=-4(\mathcal{D}_{a}\Phi)(\mathcal{D}^{a}\Phi)^{\dagger} - 4 U(|\Phi|) \nonumber\\
\textrm{with} \;\;\;\;
&U(|\Phi|)=m^{2}\Phi\Phi^{\dagger}+\lambda(\Phi\Phi^{\dagger})^{2}+\beta(\Phi\Phi^{\dagger})^{3}\,,
\label{eq:scalar}
\end{align}
\label{eqs:action}
\end{subequations}
where $\mathcal{D}=\nabla-iq A$ is the covariant derivative with respect to $A$. The scalar field $\Phi$ is of mass $m$ and charge $q$, having quartic and hexic couplings, labelled by $\lambda$ and $\beta$, respectively. The free scalar field has $\lambda=\beta=0$. Note that for $(3+1)$-dimensions, we can only take $m^{2}\geq-9/4$, to satisfy the Breitenlohner-Freedman (BF) bound.

From Eqs.~(\ref{eqs:action}), the equation of motion for this model can be derived as
\begin{subequations}
\begin{align}
R_{ab}+\frac{3}{L^{2}}g_{ab}&=2(F_{a}^{c}F_{bc}-\frac{1}{4}g_{ab}F^{cd}F_{cd})+2(\mathcal{D}_{a}\Phi)(\mathcal{D}_{b}\Phi)^{\dagger}+2(\mathcal{D}_{a}\Phi)^{\dagger}(\mathcal{D}_{b}\Phi) \nonumber\\
&+2m^{2}g_{ab}\Phi\Phi^{\dagger}+2\lambda g_{ab}(\Phi\Phi^{\dagger})^{2}+2\beta g_{ab}(\Phi\Phi^{\dagger})^{3}\,, \label{eq:einstein}
\\
\nabla_{a}F_{b}^{a}&=iq\,[(\mathcal{D}_{b}\Phi)\Phi^{\dagger}-(\mathcal{D}_{b}\Phi)^{\dagger}\Phi]\,,
\\
\mathcal{D}_a\mathcal{D}^a \Phi&=m^{2}\Phi+2\lambda(\Phi\Phi^{\dagger})\Phi+3\beta(\Phi\Phi^{\dagger})^{2}\Phi\,.
\label{eq:scalar}
\end{align}
\label{eqs:motion}
\end{subequations}
\,\;\;In the special case $\Phi=0$, it is easy to show that this model reduces to the Einstein-Maxwell theory which has been konwn to have a class of counterexamples to cosmic censorship \cite{Crisford:2017gsb}.
But when consider $\Phi\neq0$, the static solutions of Eqs.~(\ref{eqs:motion}) cannot be analytically obtained, and we will solve the equations numerically. Notice that before this, we first present the ansatz and boundary conditions for full nonlinear solutions with charged scalar.

In this work, we focus on axisymmetric and static metric, which indicates that our coordinate system can be chosen such that $\partial _t$ and $\partial _\phi$ are Killing fields. A simplest configuration of this form is the pure AdS metric in Poincar\'e coordinates
      \begin{eqnarray}\label{eqn:element}
ds^{2}=\frac{L^{2}}{z^{2}}[-dt^{2}+dr^{2}+r^{2}d\phi^{2}+dz^{2}]\,.
\end{eqnarray}
Note that in Cartesian coordinates $(r,z)$, both $r$ and $z$ are range of values in $[0,\infty)$.
To avoid this infinity, without loss of generality, we perform the coordinate transform
\begin{eqnarray}\label{eqn:trans}
z=\frac{y\sqrt{2-y^{2}}}{1-y^{2}}(1-x^{2})\,,\,\,\,\,\,\,\,\,
y=\frac{y\sqrt{2-y^{2}}}{1-y^{2}}x\sqrt{2-x^{2}}\,,
\end{eqnarray}
and adopt the polar-like coordinates $(x,y)$, where both $x$ and $y$ take values in $[0,1]$.

To obtain the desired solutions with nonzero scalar field, we generalize the pure Poincar\'e AdS by introducing the indefinite functions $Q_i$, with $i=1,2,..,7$, as functions of coordinates $x$ and $y$. The most general metric ansatz for static and axisymmetric solutions is given by
\begin{subequations}
\begin{align}
ds^{2}&=\frac{L^{2}}{(1-x^{2})^{2}}[-\frac{(1-y^{2})^{2}\,Q_{1}\,dt^{2}}{y^{2}(2-y^{2})}+\frac{4\,Q_{4}}{2-x^{2}}(dx+\frac{Q_{3}}{1-y^{2}}dy)^{2} \nonumber\\
&+\frac{4\,Q_{2}\,dy^{2}}{y^{2}(1-y^{2})^{2}(2-y^{2})^{2}}+x^{2}(2-x^{2})\,Q_{5}\,d\phi^{2}]\,,\label{eq:metric}
\end{align}
and for matter fields we take
\begin{align}
&A=L\,Q_{6}\,dt\,, \;\;\;\;\textrm{and}\;\;\;\; \Phi=(1-x^{2})^{\triangle}y^{\triangle}(2-y^{2})^{\frac{\triangle}{2}}Q_{7}\,,\label{eq:matter}
\end{align}
\label{eqs:ansatz}
\end{subequations}
with $\triangle\equiv3/2+\sqrt{9/4+m^2}$, playing the role of mass in AdS. Here we take $\triangle=2$.

Next, we will impose the appropriate boundary conditions on four boundaries, located at $x=0,1$ and $y=0, 1$, respectively.
At $x=1$, corresponding to the conformal boundary, we demand the asymptotic metric to be flat, which imposes
\begin{eqnarray}\label{eqn:boundary}
Q_{1}=Q_{2}=Q_{4}=Q_{5}=1, \;\;\;\;\;Q_{3}=0,  \;\;\;\;\textrm{and}\;\;\;\;  Q_{6}=a\,(1-y^{2})^{n},
\end{eqnarray}
where $a$ is an amplitude and $n$ denotes the fall-off behavior of asymptotic potential. At $x=0$, corresponding to the symmetry axis, regularity demands
\begin{eqnarray}\label{eqn:axis}
\frac{\partial Q_{1}}{\partial x}=\frac{\partial Q_{2}}{\partial x}=\frac{\partial Q_{4}}{\partial x}=\frac{\partial Q_{5}}{\partial x}=\frac{\partial Q_{6}}{\partial x}=0, \;\;\;\;\;Q_{3}=0,  \;\;\;\;\textrm{and}\;\;\;\;  Q_{4}=Q_{5}.
\end{eqnarray}
At the intersection of the conformal boundary with the axis of symmetry located at $r=z=0$, corresponding to the point $y=0$, we have
\begin{eqnarray}\label{eqn:intersection}
Q_{1}=Q_{2}=Q_{4}=Q_{5}=1, \;\;\;\;\;Q_{3}=0, \;\;\;\;\textrm{and}\;\;\;\; Q_{6}=a.
\end{eqnarray}
At the Poincare horizon for $n> 1$, located at $y=1$, we impose
\begin{eqnarray}\label{eqn:horizon}
Q_{1}=Q_{2}=Q_{4}=Q_{5}=1, \;\;\;\;\; Q_{3}=0, \;\;\;\;\textrm{and}\;\;\;\;  Q_{6}=0.
\end{eqnarray}

Given ansatz and boundary conditions, we could search for the full nonlinear solutions to Eqs.~(\ref{eqs:motion}) numerically and further investigate their properties, as we will discuss below.

\section{Numerical results}\label{sec3}
In this work, we use the Deturck method \cite{Headrick:2009pv} to solve the equations of motion Eqs.~(\ref{eqs:motion}), and present the smooth stationary solutions numerically.
The Deturck method, first introduced in \cite{Headrick:2009pv} and reviewed more in  \cite{Dias:2015nua}, is a effective tool to solve the Einstein equation numerically. Following the same strategy as in \cite{Dias:2015nua}, one can deform Eq.~(\ref{eq:einstein}) by adding a gauge fixing term and construct a set of solvable elliptic equations, the so-called Einstein-Deturck equation system, which reads
\begin{eqnarray}\label{eqn:equation1}
R_{ab}+\frac{3}{L^{2}}g_{ab}-\nabla_{(a}\xi_{b)}=2(F_{a}^{c}F_{bc}-\frac{1}{4}g_{ab}F^{cd}F_{cd})+2(D_{a}\Phi)(D_{b}\Phi)^{\dagger}+2(D_{a}\Phi)^{\dagger}(D_{b}\Phi)\nonumber\\
+2m^{2}g_{ab}\Phi\Phi^{\dagger}+2\lambda g_{ab}(\Phi\Phi^{\dagger})^{2}+2\beta g_{ab}(\Phi\Phi^{\dagger})^{3}\,,
\end{eqnarray}
with $\xi^{a}=[\Gamma_{cd}^{a}(g)-\Gamma_{cd}^{a}(\bar{g})]g^{cd}$. In addition, $\Gamma_{cd}^{a}(\bar{g})$ is the Levi-Civitta connection with respect to the reference metric $\bar{g}$, which is introduced to fix the residual degree of gauge freedom for the metric $g$ we intend to determine.
In this work, we choose the reference metric with $Q_{1}=Q_{2}=Q_{4}=Q_{5}=1$, $Q_{3}=0$, and $Q_{6}=a$.

Furthermore, we use the Newton-Raphson methods to iterate the elliptic equations until convergence on the finite element grid over the integrate domain $[0,1]\times[0,1]$. In particular, employing non-uniform grids allows us to refine some high gradient regions which become more important as we approach singular solutions, such that desired accuracy can be reached in smaller resource.
Note that to ensure our results correct, in this work, 200$\times$300 grid is used and the relative error below $10^{- 5}$ is required.

\subsection{Linear Solutions}\label{sec:sta}
In four-dimensional Einstein-Maxwell theory with negative cosmological constant, as amplitude $a$ increases, the electric field and spacetime curvature keep growing as well, gradually far away the pure AdS geometry. To investigate the change of spacetime geometry, one can evaluate the square of the Reimann curvature tensor $R_{\alpha\beta\gamma\delta}$, $i.e.$ the so called Kretschmann scalar $K=R_{\alpha\beta\gamma\delta}R^{\alpha\beta\gamma\delta}$.

\begin{figure}[h!]
\begin{center}
\includegraphics[height=.3\textheight,width=.37\textheight, angle =0]{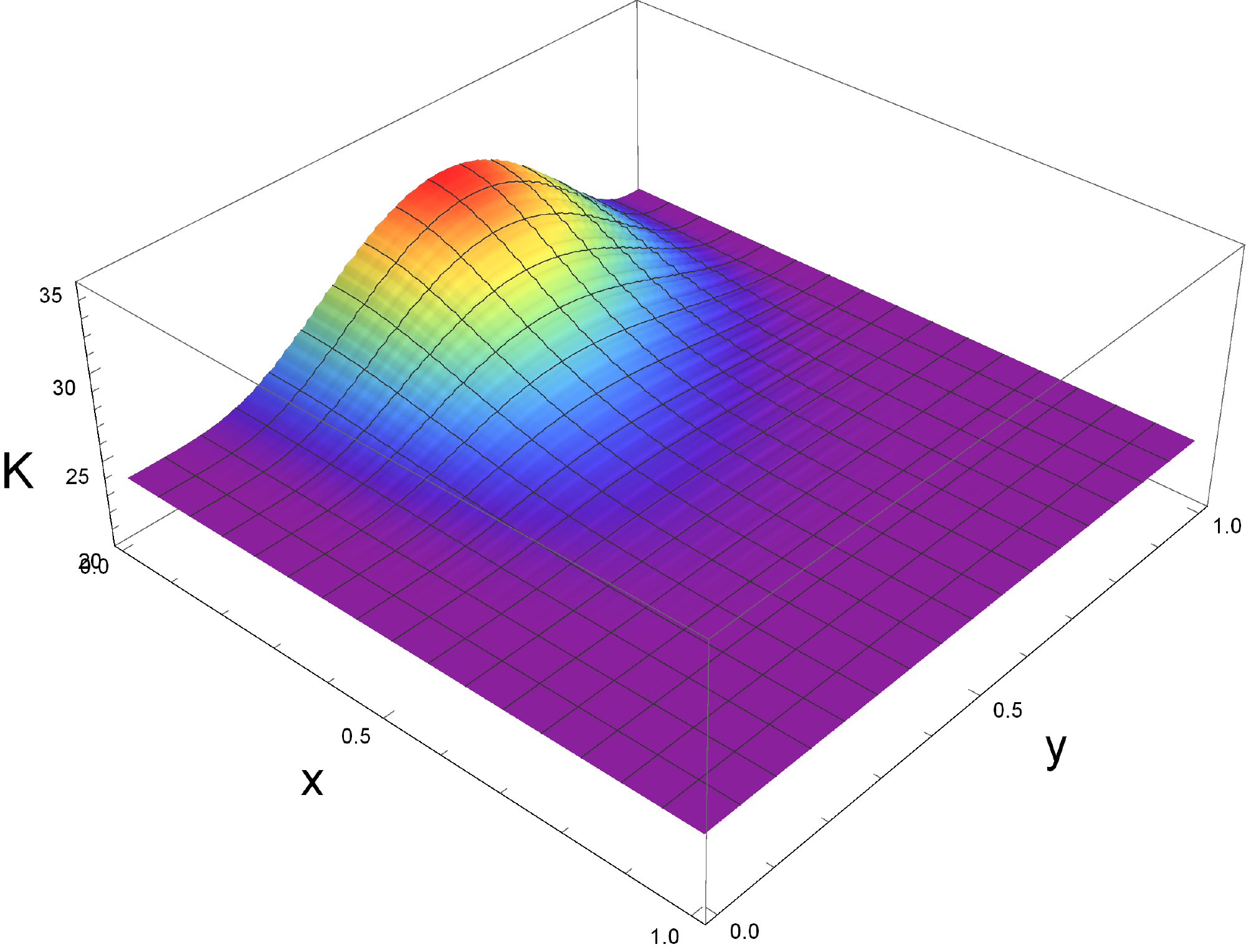}
\vspace{-1cm}
\includegraphics[height=.3\textheight,width=.37\textheight, angle =0]{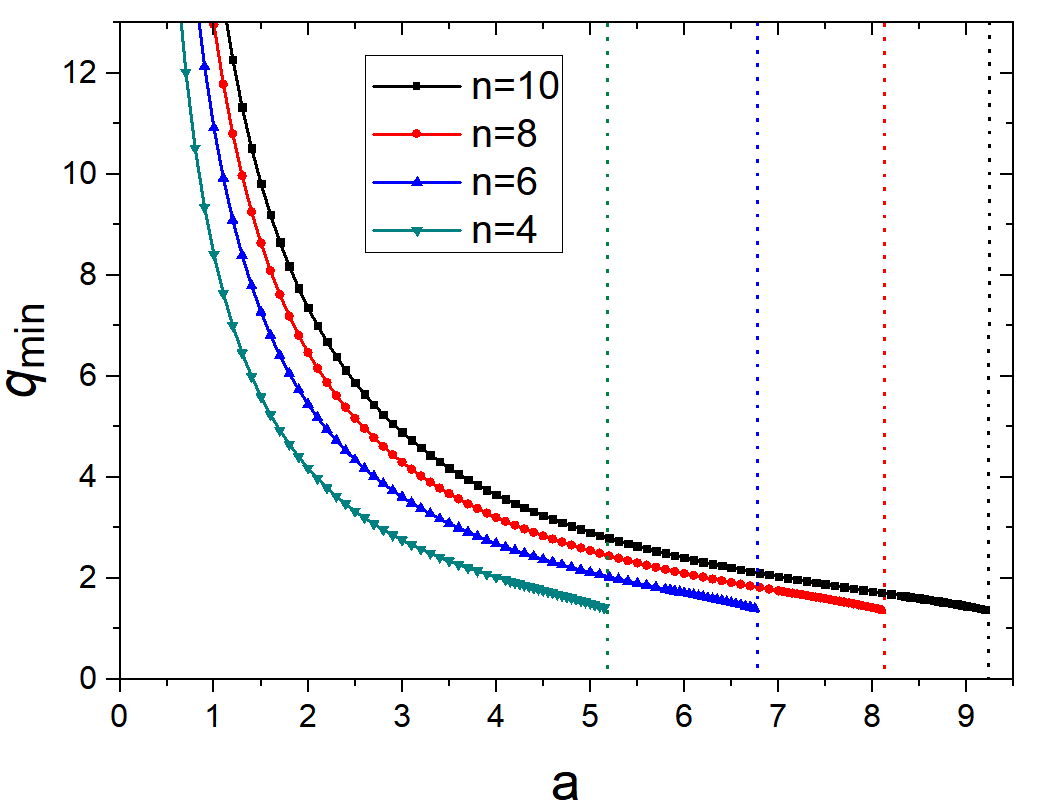}
\end{center}
\caption{\label{fig:maxk} $Left:$  The distribution of the Kretschmann scalar $K$ over the integrate domain $[0,1]\times[0,1]$, for $a=5$ and $n=8$.  $Right:$ The minimal charge $q_{min}$ against amplitude $a$ with fixed $\lambda=4$ and $\beta=6$, for various profiles $n=10, 8, 6, 4$.}

\end{figure}

We review the distribution of scalar $K$ in \cite{Horowitz:2016ezu} in $x-y$ plane for $n=8$ and $a=5$, as seen in the left panel of Fig.~\ref{fig:maxk}.
The scalar $K$ is shown to develop a maximum value at the boundary $x=0$, corresponding to the axis of symmetry. But the change in amplitude does not dramatically affect the scalar $K$ away this axis and $K$ appears to remain $K=24/L^4$, for pure AdS spacetime, as shown by the purple region.

To see if there are solutions with nonzero scalar field numerically, a simple strategy is to first investigate the stability of the Einstein-Maxwell solution \cite{Crisford:2017gsb}, and in general, the onset of instabilities can lead to new stationary solutions. According to the linear perturbation theory \cite{Dias:2015nua}, we consider a perturbative self-interacting scalar field around a fixed Einstein-Maxwell background, and then the field equation Eq.~(\ref{eq:scalar}) reduces to a linear equation
  \begin{eqnarray}\label{eq:pequation}
(\nabla_{a}\nabla^{a}-m^{2})\Phi-2\lambda(\Phi\Phi^{\dagger})\Phi-3\beta(\Phi\Phi^{\dagger})^{2}\Phi\,=q^{2}\,A_{a}A^{a}\,\Phi\,.
\end{eqnarray}
Note that in the probe limit, the scalar field along the onset of instability is so small that its higher order nonlinear terms are negligible. Hence, perturbative equation Eq.~(\ref{eq:pequation}) also reads
 \begin{eqnarray}\label{eq:ppequation}
(\nabla_{a}\nabla^{a}-m^{2})\Phi\,=q^{2}\,A_{a}A^{a}\,\Phi\,,
\end{eqnarray}
which is just the free scalar perturbative equation studied in \cite{Crisford:2017gsb}.
This interpret why the perturbative results are irrelevant to self-interacting couplings.

The numerical results for zero-modes of linear perturbations Eq.~(\ref{eq:pequation}) are presented in the right panel of Fig.~\ref{fig:maxk}, where we plot the minimal charge $q_{min}$ against amplitude $a$ for several profiles $n=4,6,8,10$. For fixed $n$ profile, the curve always ends at a maximum amplitude $a=a_{max}$, marked by the vertical dotted lines, where the electric field will diverge, and spacetime curvature grows without bound, indicating that a class of counterexample to weak cosmic censorship appear. As expected, these results are completely consistent with those for $\lambda=\beta=0$. Note that only above this zero-mode curve the Einstein-Maxwell solutions can become unstable, and the scalar field condenses and therefore the new hairy solution forms.

To see whether these counterexamples can be removed with charged self-interacting scalar field, we investigate the full nonlinear solutions derived from Eqs.~(\ref{eqs:motion}) with quartic coupling $\lambda\neq0$ and hexic coupling $\beta\neq0$, respectively.

\subsection{Full Nonlinear Solutions}\label{sec:l}
\subsubsection{Solutions with $\lambda\neq0 \,\textrm{and}\, \beta=0$}\label{sec:Rl}
To investigate the properties of hairy solution, we measure the size of scalar field $\Phi$ defined in Eqs.~(\ref{eq:matter}) by computing the expectation value of its holographic dual boundary operator $\mathcal{O}_{2}$, $\langle \mathcal{O}_2\rangle=(1-y^2)^2 U_7$, according to AdS/CFT duality. Note that we take $q=2.15>q_{min}$ to ensure scalar condensate, where $q_{min}$ is the minimal charge deduced in stability analysis.
\begin{figure}[h!]
\begin{center}
\includegraphics[height=.3\textheight,width=.37\textheight, angle =0]{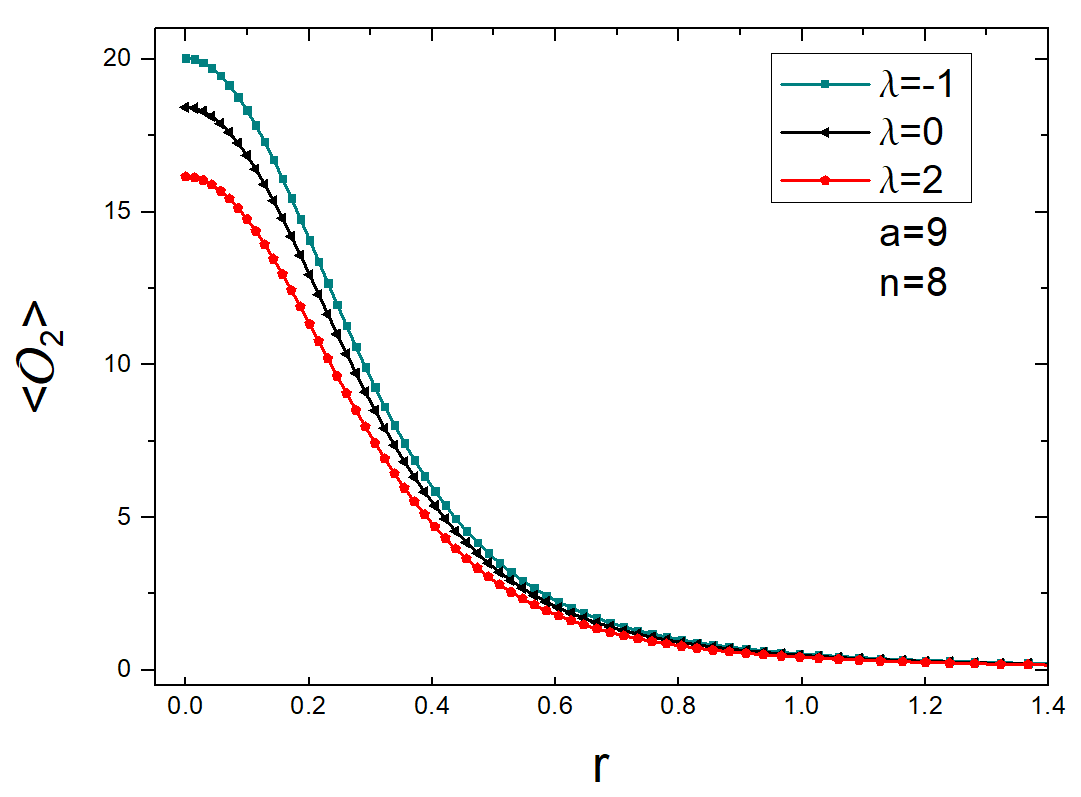}
\vspace{-1cm}
\includegraphics[height=.3\textheight,width=.37\textheight, angle =0]{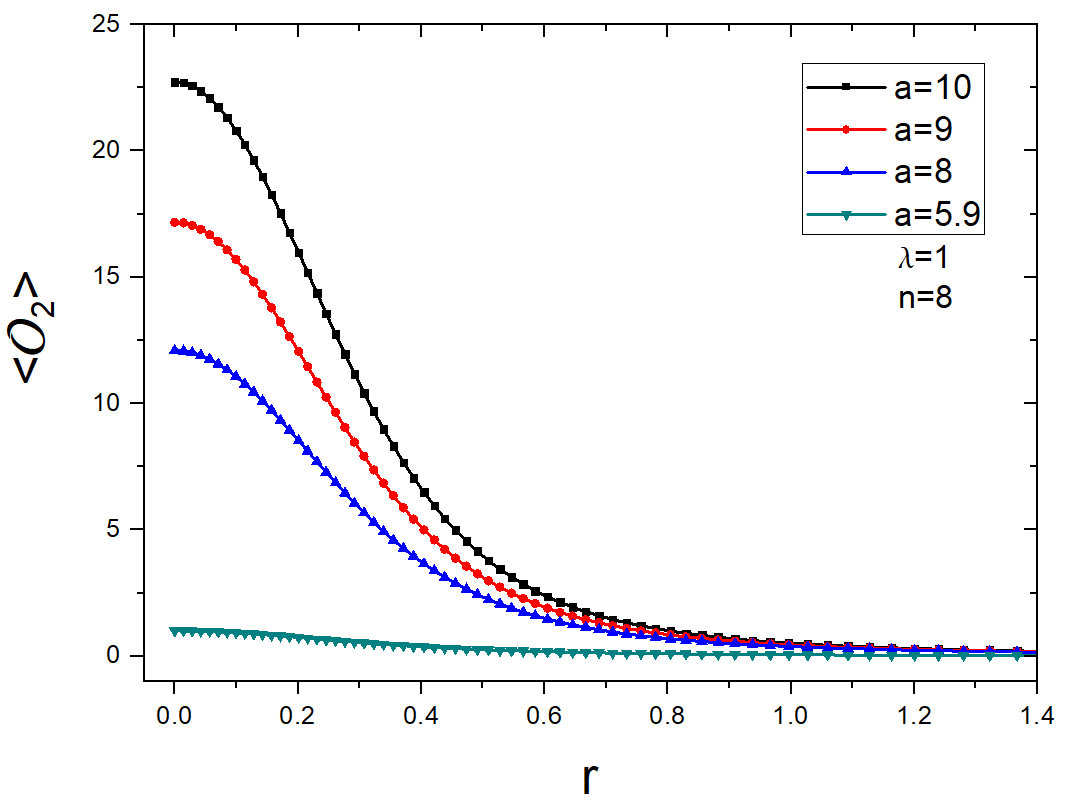}
\end{center}
\caption{ $Left:$ $\langle \mathcal{O}_2\rangle$ against $r$, for several $\lambda=-1, 0, 2$. The black solid curve is the non-interacting result for reference. $Right:$ $\langle \mathcal{O}_2\rangle$ against $r$. From top to bottom, we have $a=10, 9, 8, 7, 6$. For both panels we take $n=8$ and $q=2.15$.}
\label{fig:lcondensatevsr}
\end{figure}

In Fig.~\ref{fig:lcondensatevsr}, the left panel shows the distribution of scalar condensate $\langle \mathcal{O}_2\rangle$ at conformal boundary for fixed $a=9$ and $n=8$ for various couplings $\lambda=-1, 0, 2$, represented by green, black and red solid lines, respectively. We show that for a given value of $r$, as $\lambda$ decreases, the scalar condensate increases. Meanwhile, the right panel shows the distribution of scalar condensate $\langle \mathcal{O}_2\rangle$ with fixed $\lambda=1$. From top to bottom, these are for different $a=10, 9, 8, 7, 6$. For a given value of $r$, the scalar condensate increases with increasing $a$. In addition, from Fig.~\ref{fig:lcondensatevsr}, the scalar condensate is always maximized at the origin $r=0$ and decreases to zero at infinity. This is in agreement with the feature of holographic dual superconductor.

Form above, for fixed $q=2.15$, there indeed exist stationary smooth solutions with quartic hair at large amplitudes $a$, and the condensate is more likely to form at larger $a$. It seems the hairy solutions can exist for arbitrarily large $a$, and never be singular.

\begin{figure}[h!]
\begin{center}
\includegraphics[height=.3\textheight,width=.37\textheight, angle =0]{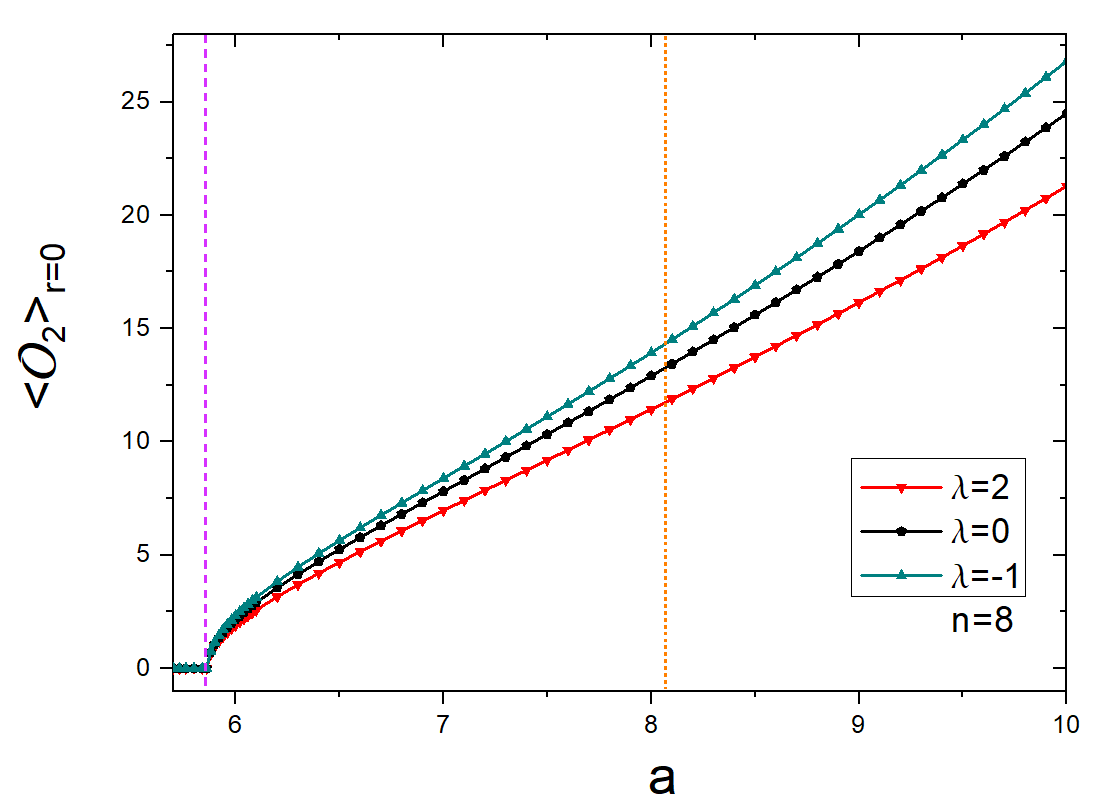}
\vspace{-1cm}
\includegraphics[height=.3\textheight,width=.37\textheight, angle =0]{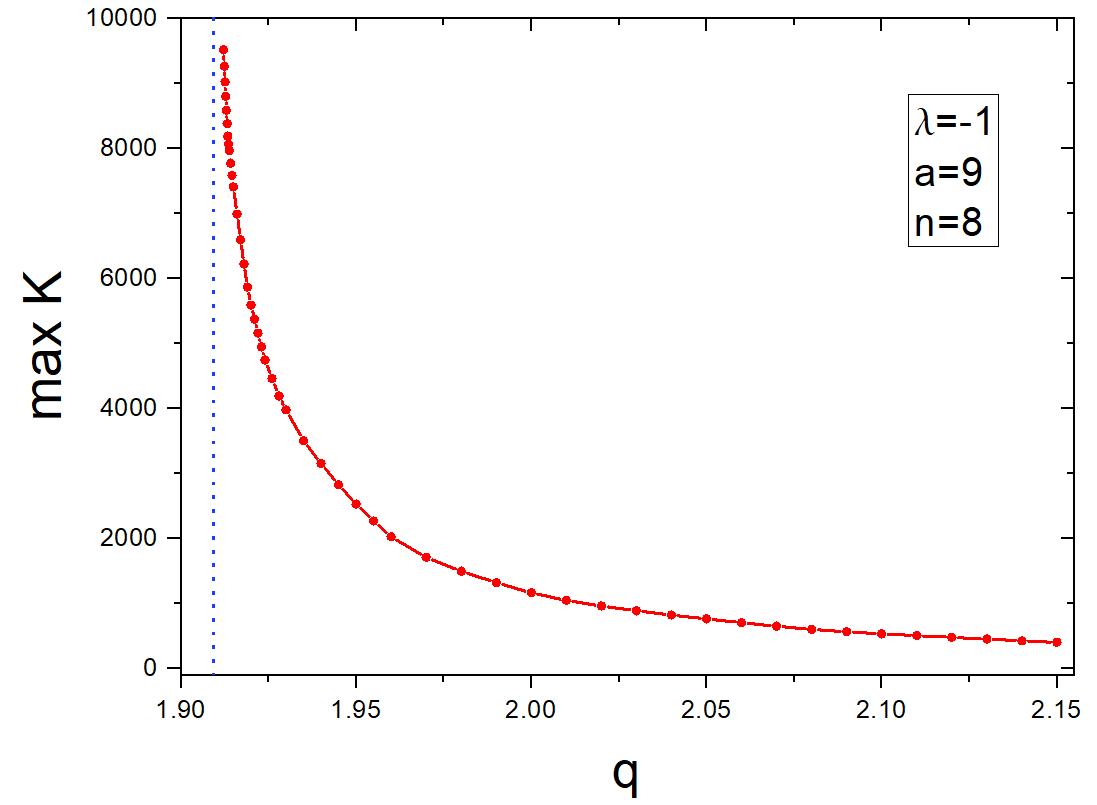}
\end{center}
\caption{
 $Left:$ Scalar condensate $\langle \mathcal{O}_2\rangle$ at $r=0$ against $a$ at given $q=2.15$ for $\lambda=-1, 0, 2$. The purple dashed vertical gridline corresponds to $a=5.86$ while the orange dotted one to $a_{max}=8.06$. $Right:$ The maximum $K$ over spacetime against scalar charge $q$, computed with $\lambda=-1$ and $a=9$. The vertical gridline marks $q=1.947$. Both panels take $n=8$.}
\label{fig:lcondensatevsa}
\end{figure}
 To test this, in the left panel of Fig.~\ref{fig:lcondensatevsa}, we present the central value of scalar condensate, $\langle \mathcal{O}_2\rangle_{r=0}$, at origin $r=0$ against amplitude $a$ with fixed $n=8$ and $q=2.15$. These curves are for $\lambda=-1$ (green), $\lambda=0$ (black) and $\lambda=1$ (red), respectively.
For each value of $\lambda$, the scalar condensate always starts to form around $a=5.86$, marked by the purple dashed vertical line, and then monotonously increases as $a$ even for $a>a_{max}$, with $a_{max}$, marked by orange dotted vertical line, being the maximum amplitude for smooth Einstein-Maxwell solution. We thus argue that the hairy solutions can exist for arbitrarily large $a$. This means that we have removed the previous counterexamples for $q=2.15$ and $n=8$.

The right panel is a plot of the maximum of Kretschmann scalar $K$ over spacetime against scalar charge $q$, with the vertical dotted line corresponding to $q_{min}=1.9123$. We set $\lambda=-1$ and $a=9$, and slowly decrease $q$ until meet an obstruction at $q=q_{min}$. The blow up as $q\rightarrow q_{min}$ indicates the appearance of a singularity. This implies that the cosmic censorship can be still violated for $a>a_{max}$ at small charge, and we therefore require the sufficiently large scalar charge $q$ to preserve cosmic censorship. Furthermore, a important question is what is the bound on charge required to preserve WCCC for arbitrarily large $a$. We will discuss this in what follows.

\begin{figure}[h!]
\begin{center}
\includegraphics[width=4in]{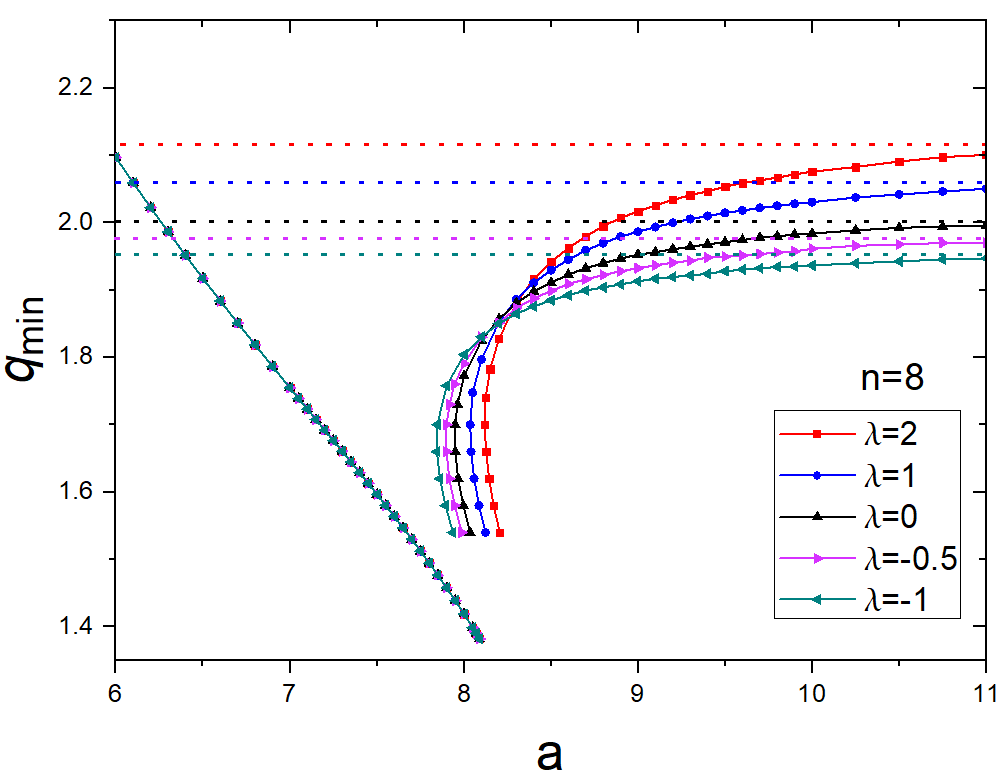}
\end{center}
\caption{
 Phase diagram of solutions with the minimal charge $q_{min}$ as a function of the amplitude $a$ with profile $n=8$, for several values of coupling parameter $\lambda$. }
\label{fig:lqminvsa}
\end{figure}
In Fig.~\ref{fig:lqminvsa} we compute the minimum charge $q_{min}$ versus various amplitude $a$ with fixed $n=8$ profile. These solid red, blue, purple and green curves correspond to $\lambda=2, 1, -0.5, -1$, respectively, with the corresponding asymptotes located at $q_{min}=2.12, 2.06, 1.975, 1.95$. The black solid line is the non-self-interacting solutions for reference. For each value of $\lambda$, there always exist two unconnected curves, one of which plotted for $a<a_{max}$, is just the onset of scalar condensate in Fig.~\ref{fig:maxk}, and the other for large $a$ is the singular curve.

For each $\lambda\neq0$, singular curve asymptotically approaches a fixed bound on charge, as seen in Fig.~\ref{fig:lqminvsa}. This is in agreement with the results for $\lambda=0$. Moveover, the effect of self-interaction also leads to manifestly differences. We show that the bound on charge presented in Fig.~\ref{fig:lqminvsa} would decrease with decreasing $\lambda$, instead of remaining the free scalar case at $q=2$. In particular, for solutions with $\lambda>0$, the bounds are located at $q>2$, but for those with $\lambda<0$ they appear at $q<2$.  That is to say, if we take $\lambda<0$, the violation would not occur even through we take $q$ slightly below $q=2$, which is not possible in free scalar case.

\subsubsection{Solutions with $\lambda=0 \,\textrm{and}\, \beta\neq0$}\label{subsec:Rll}
\begin{figure}[h!]
\begin{center}
\includegraphics[height=.3\textheight,width=.37\textheight, angle =0]{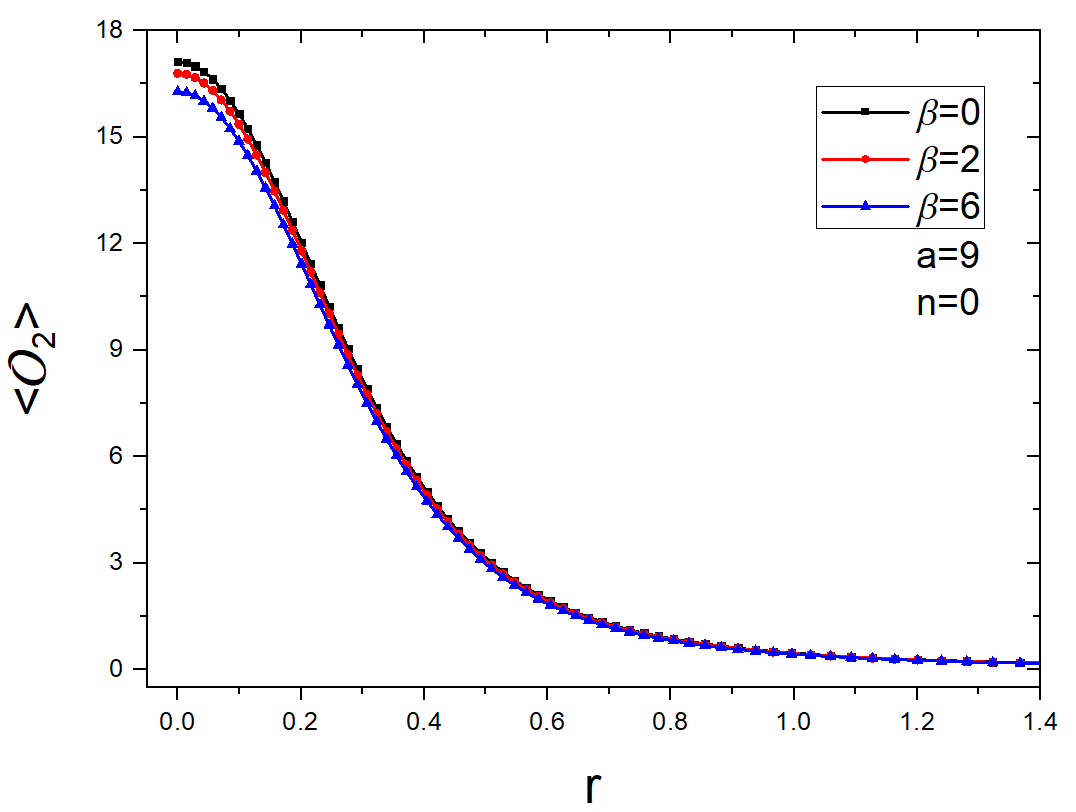}
\vspace{-1cm}
\includegraphics[height=.3\textheight,width=.37\textheight, angle =0]{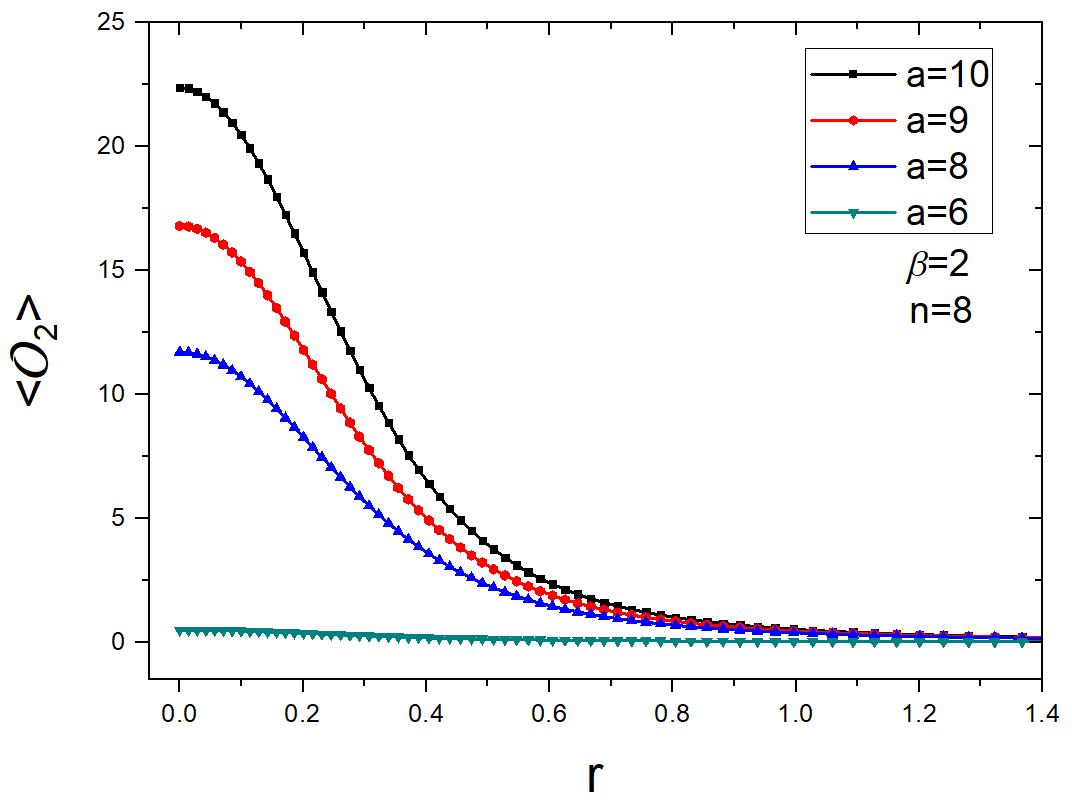}
\end{center}
\caption{$\langle \mathcal{O}_2\rangle$ as a function of $r$, for several values of $\beta$ $(left)$ and $a$ $(right)$.}
\label{fig:bcondensatevsr}
\end{figure}

To further investigate the implication of self-interaction potential, one can also analysis the full nonlinear solutions with hexic hair.
In Fig.~\ref{fig:bcondensatevsr} the condensate distribution at conformal boundary is computed with $n=8$ and $q=2.1$. Form up to bottom, these are for various $\beta=0,2,6$ (left) and for various $a=10,9,8,6$ (right), respectively. Similar to the results presented in Fig.~\ref{fig:lcondensatevsr}, the maximum condensate is always reached at origin $r=0$.
One can see that for $\beta>0$, the scalar condensate $\langle \mathcal{O}_2\rangle$ decreases as $\beta$ increases, and increases as $a$ increases. This implies that for $q=2.1$, there indeed exist stationary smooth solutions with hexic hair for large amplitudes $a$ and the hairy solutions could exist for arbitrarily large $a$.

\begin{figure}[h!]
\begin{center}
\includegraphics[height=.3\textheight,width=.37\textheight, angle =0]{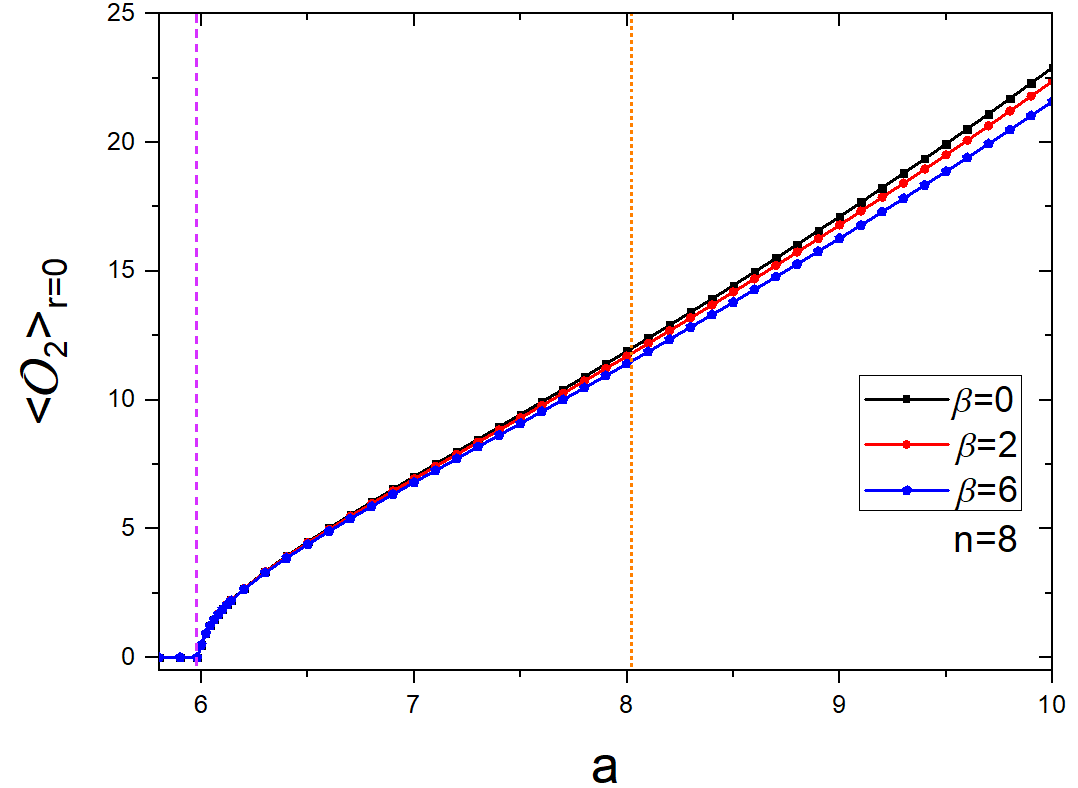}
\vspace{-1cm}
\includegraphics[height=.3\textheight,width=.37\textheight, angle =0]{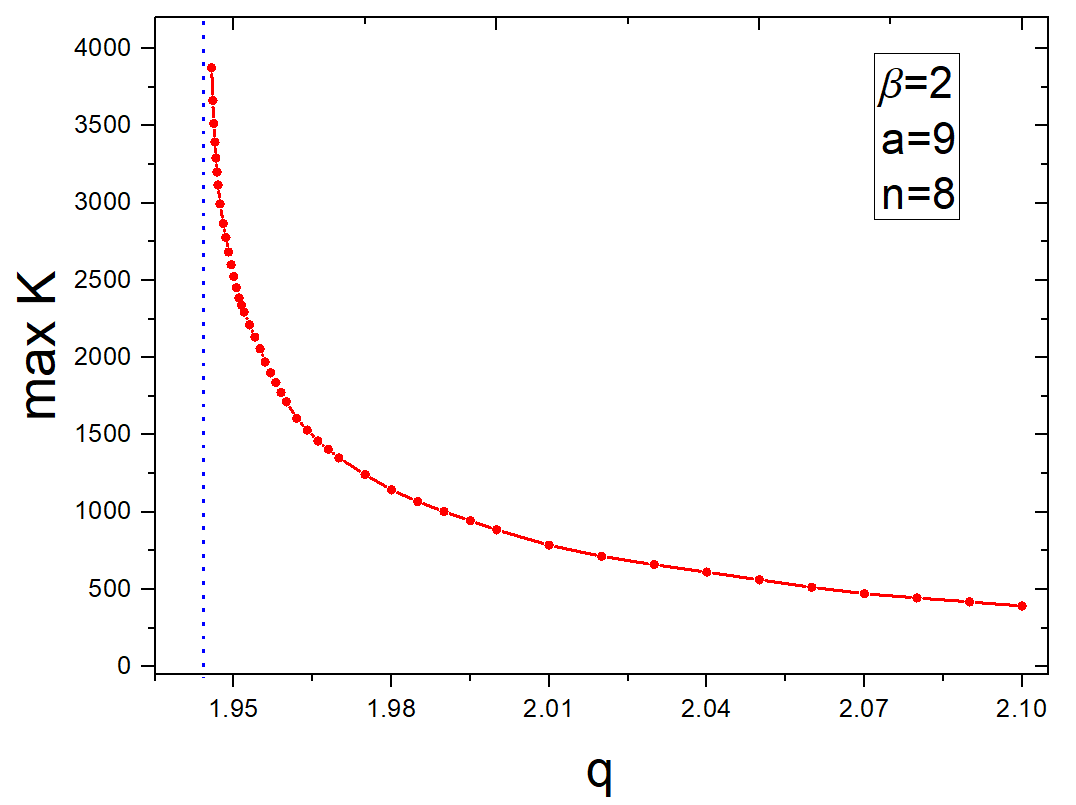}
\end{center}
\caption{$Left:$ $\langle \mathcal{O}_2\rangle_{r=0}$ against $a$ at fixed $q=2.1$, for several values of $\beta>0$ and with fixed $q=2.1$. The purple dashed vertical gridline marks $a=5.98$ while the orange dotted one $a_{max}=8.06$. $Right:$ Approach to the singular solution for fixed $\beta=2$ and $a=9$. The vertical gridline marks $q=1.945$.}
\label{fig:bcondensatevsa}
\end{figure}

 To check this, the central value of scalar condensate $\langle \mathcal{O}_2\rangle_{r=0}$ is computed with $q=2.1$, as a function of $a$. The numerical results are plotted in the left panel of Fig.~\ref{fig:bcondensatevsa}, where these curves are for various $\beta=0, 2, 6$, represented by the black, red and blue solid curves, respectively. We also mark both the onset of scalar condensate, $a=5.98$, and $a_{max}=8.06$.
For all cases for fixed $q=2.1$, the condensation curves always start with around $a=5.98$ and monotonously increase as $a$. This implies that we have removed the previous counterexamples to cosmic censorship for $q=2.1$, in the presence of hexic hair.

However, we again find that there are counterexamples for the larger amplitudes even including the scalar field, following in the quartic case. We here set $\beta=2$ and $a=9$, and it is clear from the right panel that the maximum $K$ grows without bound as we approach a minimum scalar charge, denoted by the vertical dotted line, $i.e.$ $q_{min}=1.943$. Once slightly below this minimum, a singularity will appear. This is, to avoid this singularity, the sufficiently large scalar charge must be required.

\begin{figure}[tbp]
\includegraphics[width=4in]{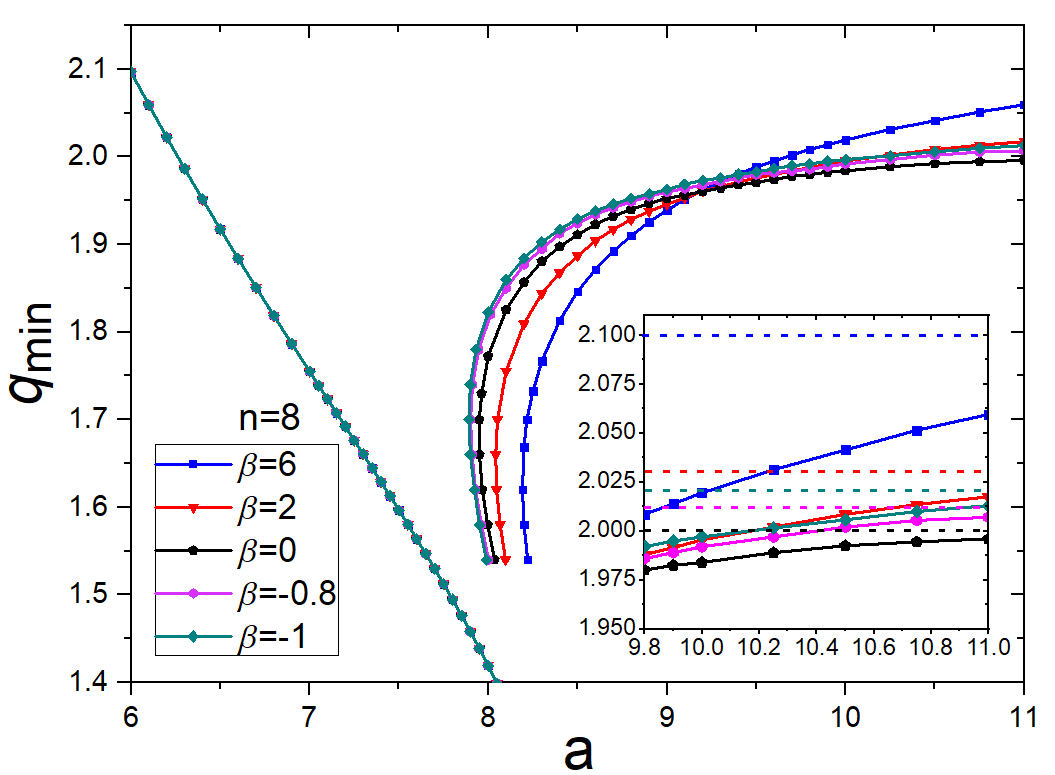}

\caption{\label{fig:bqminvsa} Phase diagram of the minimal charge $q_{min}$ as a function of the amplitude $a$, with fixed $n=8$ and for several values of $\beta=2, 6, -0.8, -1$, and the black solid line represents $\beta=0$ for reference.}

\end{figure}

One may wonder if there are also the lower bounds on charge required to preserve WCCC, as shown in the quartic case. In order to answer this, we have computed $q_{min}$ versus various $a$ in Fig.~\ref{fig:bqminvsa}.  More details are shown in the inset. These solid blue, red, purple, and green curves show $\beta=6, 2, -0.8, -1$, respectively, with the black curve for $\beta=0$ for reference. The horizonal dotted lines are just the asymptotes of these curves, marking $q_{min}=2.10, 2.03, 2.012, 2.02$.
It is clear that for each $\beta$, the singularities in hexic solutions still approach a asymptotic bound, and the value of bound increases as the absolute value of hexic coupling $\beta$ is increased. That is, for both $\beta>0$ and $\beta<0$, the bounds always are located at $q>2$, above free scalar case.

\section{Conclusion}\label{sec4}
In this work, we introduce massive charged quartic sclar field and hexic scalar field to the original Einstein-Maxwell theory in AdS which is known to have a class of counterexamples to cosmic censorship. Similar to the non-self-interacting case studied in \cite{Crisford:2017gsb}, we find that if we require the scalar charge above one bound, then these counterexamples can be removed. Besides, the self-interaction, despite does not effect the onset curve at $a<a_{max}$, indeed manifestly modifies the singular curves at larger $a$. The bounds for self-interacting theory are no longer equal to the weak gravity bound for free scalar case. For quartic solutions, the bounds are below free scalar case for $\lambda<0$, while above free scalar case for $\lambda>0$. Meanwhile, for hexic solutions, the bounds are above free scalar case for both $\beta>0$ and $\beta<0$.

In the free scalar case, the weak gravity bound is precisely the minimum charge required to
preserve cosmic censorship, implying the possible connection between WGC and WCCC. That is the main idea of \cite{Crisford:2017gsb}.
Note that though we have found that bounds on charge are below free scalar case for $\lambda<0$, this does not imply that the weak gravity-cosmic censorship connection is invalid in self-interacting theory.  It is more likely that the weak gravity bound is modified due to the presence of self-interaction. Once realize that the modified weak gravity bound on charge-to-mass ratio agrees with the one required to preserve cosmic censorship we obtained in this work, we can give a further evidence for this possible connection between these two conjecture.

We conclude with a comment about the multi-charged case. In this work, we have focused mostly on the case of a single Maxwell field, but from \cite{Horowitz:2019eum}, one can also construct a similar model of two or more Maxwell fields coupled to gravity and find similar counterexamples to cosmic censorship. Therefore, two natural problems are whether these counterexamples can be removed when we include two or more massive charged self-interacting scalar fields with different charges in the original solutions without charged scalars, and whether the minimal charge required to preserve the cosmic censorship is precisely the weak gravity bound if it exists. We will discuss these in the future work.


\section*{Acknowledgement}
We thank Shuo Sun and Hong-Bo Li for discussions.
Some computations were performed on the   shared memory system at  institute of computational physics and complex systems in Lanzhou university. This work was supported by the Fundamental Research Funds for the Central Universities (Grants No. lzujbky-2017-182).

\end{document}